\documentclass[a4paper,10pt]{article}
\usepackage[utf8]{inputenc}
\usepackage{amssymb,amsmath}
\usepackage{authblk}
\usepackage[left=0.5in, right=0.5in, top=1in, bottom=1in]{geometry}
\usepackage{hyperref}

\title{Unified statistical thermodynamics 
of quantum
gases trapped under generic power law potential in $d$ dimension and equivalence  in $d=1$}

\author{Mir Mehedi Faruk\\
Department of Theoretical Physics, University of Dhaka, Dhaka-1000\\
Theoretical Physics, Blackett Laboratory, Imperial College, London SW7 2AZ, United Kingdom
\href{mailto:me@somewhere.com}{Email: muturza3.1416@gmail.com, mehedi.faruk.mir@cern.ch} 
 }

\begin{document}
\maketitle
 
 \begin{abstract}
A unified description for the 
Bose and Fermi  gases trapped in 
an external generic power law potential $U=\sum_{i=1} ^d c_i |\frac{x_i}{a_i}|^{n_i}$
is presented using the grandpotential of the system in $d$ dimensional 
space.
The thermodynamic quantities of the quantum gases
are derived from the grand potential.
An equivalence between the trapped 
Bose and Fermi gases  is constructed in one dimension ($d=1$) using the Landen relation.
It is also found that 
the established equivalence between the ideal free Bose and Fermi gases in $d=2$ (M. H. Lee, Phys. Rev. E 55, 1518 (1997))
is lost when external potential is applied.
 \end{abstract}

\section{Introduction}
The two types of quantum gases manifest different thermodynamic behaviour
due to inherent difference of their statistical distribution\cite{pathria,huang}.  Fermi 
gas, which obeys Pauli exclusion principle  exhibit distinct characteristic such as 
zero point energy and pressure\cite{pathria,huang}
whether Bose gas condensates\cite{pathria,huang},
not obeying this principle. The thermodynamic properties of the Bose and Fermi 
gases are determined by Bose and Fermi function\cite{pathria,ziff} which have different mathematical 
structures.
But an  unified formulation  for quantum gases
was presented recently
and Lee\cite{lee,lee1,lee2} established a
remarkable equivalence between ideal free Bose and Fermi gases
in $d=2$.
The equivalence is based on a certain invariance of the polylogarithms under Euler
transformation\cite{lee3} of the fugacities.
After the inspiring work of May\cite{may} considerable attactions
are drawn to study further the equivalence between quantum gases. Point to note,
plenty of study are made to investigate the
thermodynamic properties of quantum gases under trapping potential\cite{sala,dal,toroidal,jellal}
after it was possible to create Bose-Einstein condensate in magnetically trapped Alkali gases\cite{Bradley,anderson,davis}. 
It was demonstrated in recent papers that, trapping potential can change the characteristics of quantum gases.
For instance, although there is no Bose condensation for ideal Bose gas in $d< 3$,\cite{sala,turza}
 it was found that in presence of trapping potential, Bose condensate can form in $d<3$\cite{sala,turza}.
So, it will be intruiging to check the status of this equivalence found by Lee\cite{lee}
in case of trapped system. 
\\\\ 
A lot of efforts are
made to understand dimensional dependence of 
different properties of quantum gases such  
as condensation\cite{acharyya},
\cite{acharyya2}, conductivity\cite{acharyya3},
transport properties\cite{italy}, degeneracy\cite{demarco},
polylogarithmis\cite{lee2}, q-deforemed syetm\cite{q,q1}.
Ib this report, at first the grand potential for quantum gases under generic power law potential
is determined in $d$ dimensional space.
The thermodynamic quantities are then derived 
from the grand potential. From the general expressions of the calculated 
thermodynamic quantities
we  have investigated closely
the case with $d=1$ and $n_1=2$ (harmonic potential)
and found an equivalence can be obtained
between the
thermodynamic quantities of Bose and Fermi gases.
It is also seen, the established
equivalence for ideal free quantum gases in $d=2$
disappears when a external potential is applied.
\\\\
The report is organized in the following way. 
The  grand potential  of quantum gases under generic power law potential 
is calculated 
in section 2. 
In section 3 we have presented the thermodynamic quatities
in an unified way for both types of quantum gases.
The useful landen relations are explored in section 4.
And section 5 is devoted to explore the equivalence
in $d=1$ with harmonically trapped 
quantum gases.
A discussion on the equivalence of free and trapped quantum gases is presented in section 6.
The report is concluded in section 7.

\section{Grand potential}
Considering an ideal quantum system trapped in a generic power law potential
in $d$ dimensional space with  a single particle Hamiltonian,
\begin{eqnarray}
 \epsilon (p,x_i)= bp^l + \sum_{i=1} ^d c_i |\frac{x_i}{a_i}|^{n_i}
\end{eqnarray}
Where, $b,$ $l,$ $a_i$, $c_i$, $n_i$  are all postive constants, $p$ is the momentum 
and $x_i$ is the  $i$ th component of coordinate of a particle.
Here, $c_i$, $a_i$, $n_i$ determines the depth 
and confinement power of
the potential and $l$ being the kinematic parameter. 
Now, the well known formula of density of states \cite{sala,turza},
\begin{eqnarray}
 \rho(\epsilon)&=& \int \int \frac{d^d r d^d p}{(2 \pi \hslash)^d} \delta(\epsilon - \epsilon (p,r))
\end{eqnarray}\\ 
So, from the above equation density of states is\cite{sala,turza},
\begin{eqnarray}
\rho(\epsilon)=B\frac{\Gamma(\frac{d}{l} + 1)}{\Gamma(\chi) }\epsilon^{\chi-1} 
\end{eqnarray}
where, 
\begin{eqnarray}
B=\frac{V_d C_d}{h^d a^{d/l}}\prod_{i=1} ^d \frac{\Gamma(\frac{1}{n_i} + 1)}{c_i ^{\frac{1}{n_i}}}
 \end{eqnarray}\\
Here,  $C_d=\frac{\pi^{\frac{d}{2}}}{\Gamma(d/2 + 1)} $, 
$V_d=2^d\prod_{i=1} ^d a_i$ is the volume 
of an $d$-dimensional rectangular
whose $i$-th side has length $2a_i$. 
$\Gamma(l)=\int_0 ^\infty dx x^{l-1}e^{-x}$ is the gamma function and  $\chi= \frac{d}{l} + \sum_{i=1} ^d \frac{1}{n_i}$.\\\\
The grand potential of quantum gases can be written as\cite{pathria},
\begin{eqnarray}
q=\frac{1}{a}\sum_\epsilon ln(1+azexp(-\beta \epsilon)) 
\end{eqnarray}
$\beta=\frac{1}{kT}$,
where $k$ being the Boltzmann Constant and $z=\exp(\beta \mu)$ is the fugacity, where $\mu$ being the chemical potential.
a is equal to -1 for Fermi system and +1 for Bose system.
In  experiments
with trapped gases, thermal energies far exceed the level spacing\cite{anderson}. So, using the Thomas-Fermi semiclassical 
approximation\cite{thomas} and  re-writing the previous equation,
\begin{eqnarray}
 q=q_0+\frac{1}{a}\int_0 ^\infty \ln (1+az\exp(-\beta \epsilon))\rho(\epsilon)d\epsilon
\end{eqnarray}
Here, $q_0=\frac{1}{a}\ln(1+az)$. Now finally the grand potential stands as,
\begin{eqnarray}
   q = \left\{
     \begin{array}{lr}
      q_0+ B\Gamma(\frac{d}{l}+1)(kT)^\chi f_{\chi+1}(z) &,  \text{Fermi system}\\
      q_0+ B\Gamma(\frac{d}{l}+1)(kT)^\chi g_{\chi+1}(z) &,  \text{Bose system}
     \end{array}
   \right.
\end{eqnarray}                       
Here, $g_l(z)$ and $f_l(z)$ are Bose and Fermi function respectively. Defined as
\begin{eqnarray}
 &&g_l(z)=\int_0 ^\infty \frac{x^{l-1}}{z^{-1}e^x-1}=\sum_{j=1} ^\infty\frac{z^l}{j^l} \\
 &&f_l(z)=\int_0 ^\infty \frac{x^{l-1}}{z^{-1}e^x+1}=\sum_{j=1} ^\infty (-1)^j \frac{z^l}{j^l} 
\end{eqnarray}
Now, Bose and Fermi functions can be written in terms of Polylogarithmic functions,
\begin{eqnarray}
&& Li_q(t)=g_q(t)\\
 && Li_q(-t)=-f_q(t)
\end{eqnarray}
where, $Li_q(m)$ is the polylog of $q$ and $m$. If $q\geq 1$, $Li_q(m)$ 
is analytic everywhere. It is  a real valued function if $m\in \mathbb{R}$ and $-\infty<m<1$.
A useful integral representation of polylog is
\begin{equation}
 Li_{q}(m)=\frac{1}{\Gamma(q)}\int_0 ^m [\ln(\frac{m}{\eta})]^{q-1}\frac{d\eta}{1-\eta}, 
\end{equation}
for $Re(m)<1$. To write the grand potential compactly, defining a quantity $\sigma$ as,
\begin{eqnarray}
  \sigma = \left\{
     \begin{array}{lr}
     -z &,  \text{Fermi system}\\
     z &,  \text{Bose system}
     \end{array}
   \right.
\end{eqnarray}\\
So, re writing the grand potential,
\begin{eqnarray}
   q =          q_0+ sgn(\sigma)B\Gamma(\frac{d}{l}+1)(kT)^\chi Li_{\chi+1}(\sigma) 
      \end{eqnarray}
\section{Statistical thermodynamics 
of trapped quantum
gases }
The number of particles $N$ can be obtained,
\begin{eqnarray}
&& N=z(\frac{\partial q}{\partial z})_{\beta,V}\nonumber \\
\Rightarrow &&N-N_0=  N_e=sgn(\sigma)\frac{V_d '}{{\lambda '} ^\chi}Li_\chi(\sigma)\\
\Rightarrow && \rho=\frac{N_e}{V_d '}=sgn(\sigma)\frac{1}{{\lambda '} ^d}Li_\chi(\sigma)
\end{eqnarray}
Where $V_d '$ and $\lambda'$ are defined as \cite{turza}
\begin{eqnarray}
                V_d ' &=& V_d \prod_{i=1} ^d (\frac{kT}{c_i})^{1/n_i}\Gamma(\frac{1}{n_i} + 1)\\
                \lambda'&=& \frac{h b^{\frac{1}{l} }}{\pi ^{\frac{1}{2}} (kT) ^{\frac{1}{l}}} [\frac{d/2+1}{d/l+1}]^{1/d}
               \end{eqnarray}\\
               It is noteworthy,
\begin{eqnarray}
&&\lim_{n_i\to\infty} V_d'=V_d  \\
&&\lim_{n_i\to\infty} \chi =\frac{d}{l}\\
&&\lim_{l\to 2, b\to \frac{1}{2m}} \lambda' =\lambda=\frac{h}{(2\pi mk T)^{1/2}}
\end{eqnarray}\\

           Now, the other thermodynamic quantities in case of 
           trapped system can be   calculated from grand potential as below,
          \begin{eqnarray}
 &&U=-(\frac{\partial q}{\partial \beta})_{z,V_d '}=NkT\chi\frac{Li_{\chi+1}(\sigma)}{Li_{\chi}(\sigma)}\\
 &&S=kT(\frac{\partial q}{\partial T})_{z,V_d'} -Nk\ln z +kq=Nk(\chi+1)\frac{Li_{\chi+1}(\sigma)}{Li_{\chi}(\sigma)}-\ln|\sigma|\\
 &&P=\frac{1}{\beta}(\frac{\partial q}{\partial V_d '})_{\beta,z}=NkT\frac{1}{V_d'}\frac{Li_{\chi+1}(\sigma)}{Li_{\chi}(\sigma)}\\
 &&C_V=T(\frac{\partial S}{\partial T})_{N, V_d'} =Nk[\chi(\chi+1)  \frac{Li_{\chi+1}(\sigma)}{Li_{\chi}(\sigma)} - \chi^2\frac{Li_{\chi}(\sigma)}{Li_{\chi-1}(\sigma)} ]\\
 && \kappa_T= -V_d'(\frac{\partial V_d'}{\partial P'})_{_{N,T}}=\frac{V_d'}{NkT}\frac{Li_{\chi-1}(\sigma)}{Li_{\chi}(\sigma)}        
         \end{eqnarray}\\\\
         The above expressions compactly represent the thermodynamic quantities related to trapped Bose\cite{turza,sala}
         and Fermi gas\cite{turza2,li} 
         In case of free system (all $n_i \longrightarrow \infty$) the above  quantities reduce to,
          \begin{eqnarray}
          &&\rho=sgn(\sigma)\frac{1}{{\lambda } ^d}Li_\frac{d}{2}(\sigma)\\
          &&U=NkT\frac{d}{l}\frac{Li_{\frac{d}{l}+1}(\sigma)}{Li_{\frac{d}{l}}(\sigma)}\\
 &&S=Nk(\frac{d}{l}+1)\frac{Li_{\frac{d}{l}+1}(\sigma)}{Li_{\frac{d}{l}}(\sigma)}-\log|\sigma|\\
 &&P=NkT\frac{1}{V_d}\frac{Li_{\frac{d}{l}+1}(\sigma)}{Li_{\frac{d}{l}}(\sigma)}\\
 &&C_V =Nk[\frac{d}{l}(\frac{d}{l}+1)  \frac{Li_{\frac{d}{l}+1}(\sigma)}{Li_{\frac{d}{l}}(\sigma)} - (\frac{d}{l})^2\frac{Li_{\frac{d}{l}}(\sigma)}{Li_{\frac{d}{l}-1}(\sigma)}   ]\\
 && \kappa_T=\frac{V_d}{NkT}\frac{Li_{\frac{d}{l}-1}(\sigma)}{Li_{\frac{d}{l}}(\sigma)}        
         \end{eqnarray}\\\\
So, choosing $l=2$ in case of non-relativistic quantum gas,
the Eq. (27)-(32) reduces to those in Ref.\cite{ziff} for arbitrary dimension. And with $d=3$, they
reproduce the thermodynamic quantities for free Bose
and Fermi gas\cite{pathria,huang}.

 \section{Landen Relation}
The unified formulation shows that the thermodynamic quantities 
are described by the structural properties of polylogs.  Landen\cite{lee3}
found relation between monolog and dilog, which is the 
key to make the equivalence between ideal free quantum gases\cite{lee}  as well as trapped gases.
If $x_1$ is a real number and $x_1<1$ and there exist a variable $x_2$, 
such that,
\begin{equation}
 x_2=-\frac{x_1}{1-x_1}
\end{equation}
then one  finds,
\begin{eqnarray}
&&Li_0(x_1)=-\frac{Li_0(x_2)}{Li_0(x_2)+1}\\
&& Li_1(x_1)=-Li_1(x_2)\\
 &&Li_2(x_1)=-Li_2(x_2)-\frac{1}{2}[Li(x_2)]^2
\end{eqnarray}
The proof of the above relations are included in Appendix of Ref. \cite{lee}.
These relations indicate Euler transformation\cite{lee} of $x_1$ to
$x_2$.
  \section{Application in $d=1$ for trapped gas}
Note, in both 
case of free and trapped
system the thermodynamic quantities are described by polylogs $Li_m(z)$. Now the polylogs 
are related to each other by landen relations, and the respective variables are related to each other 
by Euler transformation. 
In  free system the polylogs describing the thermodynamic system
are functions of dimension,
while in of trapped system the polylogs describing the thermodynamic system
are function
of dimension, fugacity and power law exponents. 
In trapped system the dependence of polylogs on dimension and power law
exponents are described by $\chi= \frac{d}{l} + \sum_{i=1} ^d \frac{1}{n_i}$.\\\\
As $l=2$, in case of nonrelativistic massive Boson
and choosing $d=1$, $n_1=2$ (harmonic potential),
\begin{eqnarray}
 \chi=\frac{1}{2}+\frac{1}{2}=1
\end{eqnarray}\\
If the densities are made the same,
turning our attention 
towards, density $\rho$, with $\chi=1$, we get from Eq. (27)
\begin{equation}
 \rho \lambda =Li_1(z_B)=-Li_1(-z_F)
\end{equation}
where, $z_B$ and $z_F$ denotes fugacity of Bose and Fermi gas respectively.
So, according to Eq. (33) they are related to each other by
Euler transformation. So, we can write following relation
\begin{eqnarray}
 z_F=\frac{z_B}{1-z_B}
\end{eqnarray}\\
So, the fugacities are related to each other by  Euler transformation, if we put $z_B=x_1$ and 
$z_F=-x_2$. Then, we can easily use the thermodynamic quantities
to establish the equivalence.
First turning our attention towards internal energy $U(z_B)$
with $d=1$ and $n=2$,
\begin{eqnarray}
U(z_B)&=&NkT\frac{Li_2 (z_B)}{Li_1 (z_B)} = NkT\frac{Li_2 (x_1)}{Li_1 (x_1)}=
NkT\frac{-Li_2 (x_2) -\frac{1}{2} (Li(x_2))^2 }{-Li_1 (x_2)}
=NkT[\frac{Li_2 (x_2)}{Li_1 (x_2)} +\frac{1}{2}Li_1 (x_2)]\nonumber\\
&=&U(z_F) + NkT\frac{1}{2}Li_1 (-z_F)\nonumber\\
&=&U(z_F) + NkT \rho \lambda\nonumber
\end{eqnarray}\\
Now point to note, $\rho=\frac{N_e}{V_d '}$ and $V_d ' \propto \sqrt{T}$. Also $\lambda \propto \frac{1}{ \sqrt{T}}$
So, obviously the second term is temperature independent.
As it turns out, the second term exactly corresponds to ground state energy \cite{turza2}
just in the case of ideal free quantum gases\cite{lee}.
Hence, it can be concluded if the two reduced densities are the same,
the fugacities are related by Euler transformation and as a result internal energies of Bose and Fermi gases 
only differ by the ground state energy of the Fermi gas only.
So, denoting ground state energy by $U_0$, we can rewrite,
\begin{eqnarray}
U(z_B)=U(z_F)-U_0
\end{eqnarray}\\
Since,
pressure and energy are related by  $PV_d '=\frac{E}{\chi}$, from the help of
Eq. (40), one can get
\begin{eqnarray}
 P(z_B)=P(z_F)-P_0
\end{eqnarray}
Where, $P_0$ denotes ground state pressure of Fermi gas\cite{turza}.
Now turning our attention towards entropy,
\begin{eqnarray}
 S(z_B)&=&Nk[2\frac{Li_2(z_B)}{Li_1(z_B)} -log(z_B)]
 =Nk[2\frac{Li_2(x_1)}{Li_1(x_1)} -log(x_1)]
=Nk[2\frac{Li_2(x_2)+\frac{1}{2}[Li_1(x_2)]^2}{Li_1(x_2)} -log(\frac{-x_2}{1+x_2}) ]\nonumber \\
&=& Nk[ 2\frac{Li_2(x_2)}{Li_1(x_2)} + Li_1(x_2) - log(-x_2) +log(1+x_2)
= Nk[ 2\frac{Li_2(x_2)}{Li_1(x_2)}  - log(-x_2)]\nonumber \\
&=& Nk[ 2\frac{Li_2(-z_F)}{Li_1(-z_F)}  - log(z_F)]
=S(z_F)
\end{eqnarray}
Here, we have used the identity $log(1+x)=- Li_1(x)$.
Also, it is clear that entropy remain exactly same for two types of
quantum gases in this case. 
Now, from the equation of specific heat,
\begin{eqnarray}
 C_V(z_B)&=&Nk[2\frac{Li_2(z_B)}{Li_1(z_B)} -  \frac{Li_1(z_B)}{Li_0(z_B)}]\nonumber \\
 &=&Nk[2\frac{Li_2(x_1)}{Li_1(x_1)} -  \frac{Li_1(x_1)}{Li_0(x_1)}]
 =Nk[2\frac{Li_2(x_1)}{Li_1(x_1)} -  \frac{Li_1(x_1)}{Li_0(x_1)}]\nonumber \\
 &=& Nk[2\frac{Li_2(x_2)+\frac{1}{2}[Li_1(x_2)]^2}{Li_1(x_2)} -  \frac{Li_1(x_2)[1+Li_0(x_2)]}{Li_0(x_2)}]\nonumber \\
 &=&Nk[2\frac{Li_2(x_2)}{Li_1(x_2)} -  \frac{Li_1(x_2)}{Li_0(x_2)}] 
 = Nk[2\frac{Li_2(-z_F)}{Li_1(-z_F)} -  \frac{Li_1(-z_F)}{Li_0(-z_F)}]
 =C_V(z_F)
 \end{eqnarray}
This type of result is previously found by  May\cite{may}  for free quantum gases in two dimensional space.   
 In case of the isothermal 
compressibilty, 
\begin{eqnarray}
 \kappa_T(z_B)=\frac{V_d'}{NkT}\frac{Li_{0}(z_B)}{Li_{1}(z_B)}=(1+z_F)\kappa_T(z_F)
\end{eqnarray}
So, the isothermal compressibilty
 are not equivalent at all temperatures. 
 If $z_F\longrightarrow \infty$  (i.e $z_B \longrightarrow 1$), $\kappa_T (z_F )\propto \frac{1}{log(z_F)}$ but
$\kappa_T (z_B )\propto \frac{z_F}{log(z_F)}$.
  The latter diverges while the former vanishes.
  But if $z_F\longrightarrow 0$ (in the classical limit)
 the two of course become equivalent.
 The same conclusion
applies to the number fluctuation. \\\\
Finally, the reasons behind this remarkable equivalence
can be found from the grand potential $Q$.
Rewriting $Q$,
\begin{eqnarray}
\log Q (z_B)&=&  \frac{V_d '}{\lambda } Li_{2}(x_B)  
            = \frac{V_d '}{\lambda } Li_{2}(x_1)  
            = \frac{V_d '}{\lambda } (-Li_{2}(x_2)-\frac{1}{2}[Li_1(x_2)]^2 )\nonumber\\  
            & = &\frac{V_d '}{\lambda } (-Li_{2}(-z_F)-\frac{1}{2}[Li_1(z_F)]^2 )  \nonumber\\
            & = & \log Q(z_F)-\frac{V_d '}{2} \rho^2 \lambda^3  
            \end{eqnarray}\\
With careful inspection it can be seen from Eq. (16) - (18),
the second term ib Eq. (45) is linear in $\beta$. So, the grand partition function
of the two systems are related to each other by a term linear in $\beta$.
As all the thermodynamic quantities are basically dervied from grand potential,
thus we are able to make such connection for all the thermodynamic
quantities. So, when we take first derivative of grand potential with respect 
to $\beta$, the obtained thermodynamic quantity internal energy of Bose and Fermi 
system only differ by  a constant ($i.e.$ the ground state energy) which is 
independent of $\beta$. And when we take
the second derivative of grand potential with respect to $\beta$, the derived thermodynamic 
quantity specific heat are equal to each other. \\
\section{Discussion}
In this paper, we have seen once again if the fugacities of Bose and Fermi gas 
are related by Euler transformation an equivalence relation can be establised 
between the two types of quantum gases. One can check the status of the equivalence relation in $d=2$ \cite{lee}
for trapped quantum gases.
Now re-writing the equation of reduced density (Eq. 27) with  $l=2$,
\begin{eqnarray}
 \rho \lambda^2 = Li_{\frac{d}{2}+\sum_{i}\frac{1}{n_i}} (z_B) =- Li_{\frac{d}{2}+\sum_{i}\frac{1}{n_i}} (-z_F)  
\end{eqnarray}
Now, choosing $d=2$, the above expression
reduces to,
\begin{eqnarray}
 \rho \lambda^2 = Li_{_{1+\sum_{i=1,2}\frac{1}{n_i}}} (z_B) = -Li_{_{1+\sum_{i=1,2}\frac{1}{n_i}}} (-z_F) \nonumber 
\end{eqnarray}\\
From Eq. (35) one can see the Euler transformation type relation between fugacities are possible only for monologs.
So, as it stands from the above equation the Euler transformation type relation between fugacities are
possible if and only if $\sum_{i=1,2}\frac{1}{n_i}=0$.
Now as,  $n_1,n_2>0$
this criterion is possible if and only if $n_1\longrightarrow \infty$ and $n_2\longrightarrow \infty$, which 
is basically the condition for free system \cite{sala,jellal,turza,turza2}. So, the equivalence relation between
the quantum gases is possible in two dimensional space only for the free system.
This phenomenon is due to the fact,
that trapped Bose gas actually condensates in $d=2$ with any trapping potential or
in more general, BEC is possible if and only if $\chi>1$\cite{sala,jellal,turza}.
So, the equivalence relation between the Bose and Fermi gases are possible 
where BEC could not take place. Now in case of $d=1$, Eq. (46) becomes,
\begin{equation}
 \rho \lambda = Li_{\frac{1}{2}+\frac{1}{n_1}} (z_B) =- Li_{\frac{1}{2}+\frac{1}{n_1}} (-z_F)  
\end{equation}
Again Eq. (35) suggests 
Euler transformation type relation between fugacities are possible  in $d=1$
if and only if quantum gases are trapped in harmonic potential $(n_1=2)$.
\section{Conclusion}
From the unified statistical thermodynamics 
of quantum
gases trapped under generic power law potential in $d$ dimension, 
a case is shown with  $d=1$ where,  Bose and Fermi gases can be treated as equivalent. This is possible only when
the quantum gas is trapped in harmonic potential. It will be interesting to check 
the effect of interaction on this equivalence as well as to do the whole calculation with relativistic hamiltonian.
\section{Acknowledgement}
I would like to thank Fathema Farjana and Dr. Jens Roder
for their effort to help me present the manuscript.

  \end{document}